\documentclass{aa}
\usepackage{graphicx}
\usepackage{amsmath}   
\usepackage{amssymb}
\usepackage{times}

\def\eq#1{{Eq.~(\ref{#1})}}

\begin{document}

\title{Distinguishing AGN from starbursts as the origin of double peaked Lyman-Alpha Emitters in the reionization era}
\titlerunning{Double-peaked LAEs during reionization}

 \author{Hamsa Padmanabhan
          \inst{1}
          \and
          Abraham Loeb\inst{2}
          }

   \institute{D\'epartement de Physique Th\'eorique, Universit\'e de Gen\`eve \\
24 quai Ernest-Ansermet, CH 1211 Gen\`eve 4, Switzerland
   \\
              \email{hamsa.padmanabhan@unige.ch}
         \and
            Astronomy department, Harvard University \\
60 Garden Street, Cambridge, MA 02138, USA \\
             \email{aloeb@cfa.harvard.edu}            
             }

   \date{}

\abstract
{We discuss the possible origin of the double-peaked profiles recently observed in  Lyman-Alpha Emitters (LAEs) at the epoch of reionization ($z \gtrsim 6.5$) from obscured active galactic nuclei (AGN). 
Combining the extent of the Lyman-$\alpha$ near-zones  estimated from the blue peak velocity offset in these galaxies,  with the ionizing emissivity of quasars at $z \gtrsim 6$, we forecast the  intrinsic UV and  X-ray luminosities of the AGN needed to give rise to their double-peaked profiles. We also estimate the extent of the obscuration of the AGN by comparing their luminosities to those of similar quasar samples at these epochs. Future X-ray and radio observations, as well as those with the \textit{James Webb Space Telescope}, will be valuable tools to test the AGN contribution to the intergalactic-scale ionization zones of high-redshift LAEs.}


\maketitle

\section{Introduction}

The epoch of reionization (EoR), when intergalactic hydrogen transitioned from being neutral to ionized, is within the reach of current galaxy surveys \citep[e.g.,][]{loeb2013}. The Lyman-$\alpha$ line (rest wavelength 1215 \AA) is an important probe of reionization since both its strength and lineshape are sensitive to the amount of neutral hydrogen \citep[for a review, see e.g.,][]{djikstra2014}. Surveys of Lyman-Alpha Emitting galaxies (LAEs) at $z \sim 6$ can potentially constrain the timing, topology and sources of reionization \citep[e.g.,][]{stark2010, fontana2010}, as well as the properties of the ionized bubbles \citep[e.g.,][]{mason2020}. Several LAEs detected at high redshifts \citep{ouchi2010,kashikawa2006, hu2010, konno2014} close to the EoR  have been used to measure the average neutral fraction of hydrogen. The observed rapid decline in the fraction of LAEs  beyond $z \sim 6$ \citep{konno2014, sadoun2017, matthee2015, bagley2017,ota2017,shibuya2018} compared to continuum selected galaxies has been used to place constraints on the timing and characteristics of reionization \citep[e.g.,][]{dayal2009, bolton2013, tilvi2014, mesinger2015,  choudhury2015, weinberger2018,  mason2018, santos2016, yang2020, whitler2020}, though the modelling is complicated since it may also involve changes in the intrinsic properties of the LAEs with redshift \citep[e.g.,][]{stark2015, sobral2017, hassan2020}. However,  the most luminous LAEs show less evolution \citep{santos2016, ota2017, zheng2017, konno2018}  over the $z \sim 5-7$ range.
 
 Recently, some extremely luminous LAEs have been detected at $z \sim 6.6$, a few of which show evidence for double-peaked line profiles.  This is rare among high-redshift galaxies \citep[e.g.,][]{gronke2020}, since the increasing neutral fraction in the intergalactic medium \citep{hu2010} almost always leads to the resonant absorption of the blue wing of the line. Its presence, therefore, indicates that the LAE resides in a highly ionized HII region which allows the blue wing to emerge  \citep[e.g,][]{djikstra2014}, and can constrain the minimum size of the ionized region in which the galaxy resides \citep[e.g.,][]{matthee2018, mason2020}. Double-peaked LAEs can also be used to  constrain the escape fraction of ionizing photons from the parent galaxy, which was shown \citep{verhamme2015, izotov2018} to be tightly correlated with the velocity separation between the red and blue peaks.  One of the most studied double-peaked LAEs at $z \gtrsim 6.5$  is  the galaxy labeled COLA-1, located in the COSMOS field \citep{hu2016, matthee2018} at $z \sim 6.593$. Recently, two other LAEs have also been reported to have double-peaked profiles: NEPLA-4, at $z \sim 6.54$ \citep{songaila2018}, and A370p\_z1 in the parallel Frontier Field of Abell 370 \citep{meyer2020}.

At low redshifts, almost all the luminous LAEs (with $\log [L_{\alpha}/\text{ergs/s}] > 43.4)$ are found to be associated with Active Galactic Nuclei \citep[AGN;][though there are also  luminous Lyman-$\alpha$ galaxies that show no evidence for AGN, e.g. \citet{marques2020}]{konno2016}. It is possible that AGN activity makes a contribution  to the ionization region and thus the observed double peak of the above high-redshift LAEs \citep{barkana2003, matthee2018, meyer2020,pellicia2020} as well.\footnote{\citet{bosman2020} recently  reported the presence of a double peaked profile in Aerith B, a $z \sim 5.8$ Lyman-Break selected galaxy, located in the near-zone of a quasar.} Unobscured AGN are ruled out in the above scenario due to the narrowness of the observed Lyman-$\alpha$ line. However,  obscured AGN \citep[for a review, see, e.g. ][]{hickox2018} are surrounded by an edge-on torus along our line-of-sight, which blocks their broad emission lines. While the AGN's own broad Lyman-$\alpha$ line is removed from view by the obscuration \citep[e.g.,][]{baek2013}, the UV radiation is able to escape in the perpendicular direction along the torus symmetry axis, creating two ionization cones and thus the observed double-peak of the galaxy's Lyman-$\alpha$ emission. Distinguishing obscured AGN from starbursts  at high redshifts \citep[e.g.,][]{pallottini2015, pacucci2017,matthee2020} is  important for calibrating  the as-yet uncertain contribution of AGN to reionization \citep[e.g.,][]{madau2015, hassan2018, mitra2018}.

In this paper, we point out observable signatures of the double-peaked profiles in high-redshift LAEs being the result of obscured AGN, using COLA-1 as an example. If the ionizing emissivity of the quasar is sufficiently high, double-peaked emission can result due to the extent of the quasar's Lyman-$\alpha$ near zone \citep{mason2020}. We forecast the expected spectra and intrinsic X-ray emission of the AGN in the relevant energy bands, as well as the extent of its obscuration.

\section{Conditions for double-peaked profiles from AGN}

We start with the formalism of \citet{bolton2007} for the ionized near-zones  from the spectra of $z \sim 6$ quasars. It was found that the maximum radius (in proper Mpc) of the Lyman-alpha near-zone, i.e. the region around which the ionized bubble becomes optically thick to Lyman-$\alpha$ photons,  is given by:
\begin{eqnarray}
R_{\alpha, \rm max} &=& 3.14  \left(\frac{N_{\rm ion}}{2 \times 10^{57} s^{-1}}\right)^{1/2} \left[\frac{(\alpha_{\nu})^{-1} (\alpha_{\nu} + 3)}{3}\right]^{-1/2} \nonumber \\
&& \times \left(\frac{1 + z_{\alpha}}{7}\right)^{-9/4} \text{pMpc} \, ,
\label{rmax}
\end{eqnarray}
for a quasar with spectral index $\alpha_{\nu}$ and ionizing emissivity $N_{\rm ion}$ (in s$^{-1}$), where $z_{\alpha}$ is the redshift of the near-zone boundary.\footnote{For simplicity, we do not distinguish between the redshift of the source and $z_{\alpha}$ in the calculation that follows.} The above expression assumes that the density of the neutral gas is equal to the cosmic mean at the near-zone boundary, and the IGM temperature in the vicinity of the quasar is $2 \times 10^4$ K, a typical value at these epochs \citep[e.g.,][]{bolton2010}. The edge of the near-zone is defined by the last pixel at which the spectrum drops below the normalized flux limit of 10\%, corresponding to an optical depth detection limit of $\tau = 2.3$.

We combine the above expression with the findings of \citet{mason2020}, who connect the observed blue velocity offset $\Delta v_{\alpha}$ in double-peaked LAEs to the minimum extent of the underlying Lyman-$\alpha$ near-zone, $R_{\alpha, \text{min}}$, {following their assumption of a single source residing at the centre of a large, highly ionized bubble (in contrast to numerous mildly-ionized bubbles and/or neutral IGM scenarios also considered in \citet{matthee2018}).}  This is given by $R_{\alpha, \text{min}} = |\Delta v_{\alpha}|/H(z_s)$ where $H(z_s)$ is the Hubble parameter at the source redshift $z_s$,  since the photons must travel a minimum distance of $R_{\alpha, \text{min}}$ so that they redshift out of resonance while still located within the proximity zone.  In the case of COLA-1, the observed $\Delta v_{\alpha} = -250$  km/s leads to
$R_{\alpha, \rm{COLA-1}} = 0.31$ pMpc.

Equating the right-hand side of \eq{rmax} to $R_{\alpha, \rm{COLA-1}}$, we can estimate the quasar ionizing emissivity that is required for the  observation of the double peak. To do this, we need to invoke a quasar spectral shape.  We use the parametrization of \citet{shen2020} in which 
the quasar spectrum has  a power law having slope $\alpha_{\nu} = -1.70$ in the 600  \AA\ to 912 \AA\ regime \citep{lusso2015}, and a power law with an exponential cutoff for wavelengths $<$ 50 \AA\ :

\begin{equation}
L_{\nu} = 
\begin{cases}
L_{\rm H} \left({\nu/\nu_{\rm H}}\right)^{\alpha_{\nu}}; \ & 600 \text{\AA} < \lambda < 912 \text{\AA}
\\
A E^{1- \Gamma} \exp\left(-E/E_c\right) ;  & \lambda < \ 50 \text{\AA}
\end{cases}
\label{lnudoublepower}
\end{equation}
where $E_c = 300$ keV; $E = h \nu$ with $h$ being Planck's constant, $\alpha_{\nu} = -1.7$, $\Gamma = 1.9$. Here, $\nu_{\rm H}$ is the ionization frequency of hydrogen  (corresponding to 912 \AA) and the 600 \AA\  and 50 \AA\ parts of the spectrum are connected directly due to the lack of sufficient observational data. \footnote{Recent work \citep{wang2020} may indicate evidence for an increase in $\Gamma$ at high redshifts, $z \sim 6$.} {In the above equation, $L_{\rm H}$ is the luminosity of the quasar at the hydrogen ionizing limit $\nu_{\rm H}$. $A$ is the normalization of the luminosity in the X-ray regime, fixed by scaling the spectrum to the parametrization in \citet{shen2020}.}

\begin{figure}
\begin{center}
\includegraphics[width = \columnwidth]{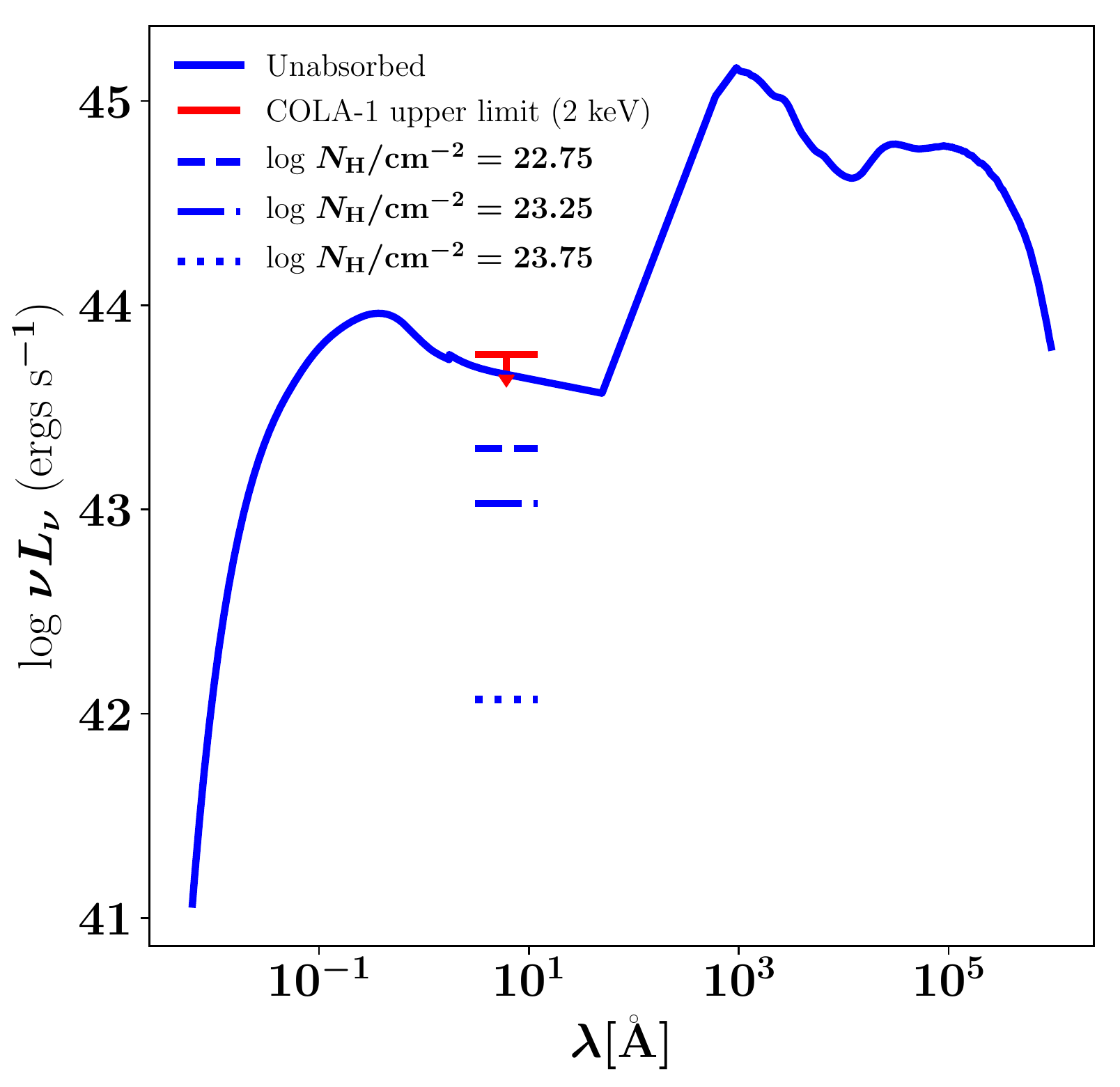}
\end{center}
\caption{SED of an AGN consistent with the observed double peak emission observed in COLA-1. The red downward arrow shows the measured upper limit in X-rays (Jorryt Matthee and Jo\~ao Calhau, private communication), rescaled to 2 keV. The blue horizontal lines indicate the suppression in the flux at 5 keV for various column densities, $N_{\rm H} = 10^{22.75}, 10^{23.25}$ and $10^{23.75}$ cm$^{-2}$, obscuring the AGN.}
\label{fig:cola1}
\end{figure}

Using $\alpha_{\nu} = -1.70$ in \eq{rmax}, we find that the required quasar emissivity corresponding to  $R_{\alpha, \rm max, COLA-1} = 0.31$ pMpc is
\begin{equation}
N_{\rm{ion}, \rm{COLA-1}} = 2.60 \times 10^{55} \ \text{s}^{-1}
\label{nioncola1}
\end{equation}
This ionizing emissivity can be used to normalize the spectral shape, by using the relation:
\begin{equation}
N_{\rm ion} = \int_{\nu_{\rm H}}^{\infty} \frac{L_{\nu}}{h \nu} d \nu  \, ,
\label{nionlnu}
\end{equation}
with $L_{\nu}$ given by\footnote{In practice, the contribution from the X-ray section of \eq{lnudoublepower} is several orders of magnitude  lower than that of the 600 - 912 \AA\ portion.} \eq{lnudoublepower}.
Using \eq{nionlnu} with the value of COLA-1's ionizing emissivity to normalize \eq{lnudoublepower},  we can  plot the flux spectrum of the AGN in the UV to IR regime, shown  as the blue line in Fig. \ref{fig:cola1}. We find that this spectrum corresponds to a flux density at 4500 \AA\ of log $L_{\nu, 4500} = 45.08$ ergs/s/Hz, or a UV magnitude of $M_{1500} = -23.2$ for the quasar. {Note that this intrinsic magnitude is brighter than the UV magnitude estimated from the COLA-1 galaxy itself \citep[$M_{\rm UV} = -21.6$;][]{matthee2018}. The suppression in the observed magnitude to below -21.6 is consistent with the required dust obscuration of the quasar that leads to extinction.}

This can now be used to normalize the  the X-ray portion of \eq{lnudoublepower}, whose integral between 0.5 keV and 7 keV gives the expected luminosity of the quasar in the X-ray regime.\footnote{To normalize the fIux at 0.5 keV, we use the $\alpha_{\rm ox}$ parameter corresponding to 2500 \AA\ \citep{steffen2006} and defined by:
\begin{equation}
 \alpha_{\rm ox} = \frac{\log(L_{0.5 {\rm keV}}/L_{2500 \text{\AA}})}{\log (\nu_X/\nu_{\rm UV})}
 \label{alphadef}
\end{equation}
  where $\nu_X$ and $\nu_{\rm UV}$ are frequencies corresponding to 0.5 keV and 2500 \AA\ respectively \citep[e.g.,][]{elvis2010}.} For COLA-1, we find $L_{0.5 - 7 \text{keV}} = 10^{44.17}$ ergs/s using the above procedure, which is  just below the measured upper limit of $10^{44.3}$ ergs/s (Jorryt Matthee \& Jo\~ao Calhau, private communication). This indicates that by slightly improving the sensitivity of the X-ray observations, it may be possible to confirm or rule out the AGN scenario for COLA-1, unless the obscuring torus has a very large gas column that suppresses the X-ray emission along the line-of-sight considerably.

Using the bolometric correction of \citet{runnoe2012}, this intrinsic X-ray luminosity corresponds to a the bolometric luminosity $L_{\rm Bol} = 10^{45.9}$ erg/s (assuming a radio quiet quasar) or  $L_{\rm Bol} = 10^{46.0}$ erg/s (assuming a radio loud quasar), both of which indicate a central black hole of mass $M_{\rm BH} \sim 10^8 M_{\odot}$ accreting at the Eddington limit.

The other two LAEs at $z \sim 6$, NEPLA-4 and A370p\_z1 can be analyzed by a similar procedure, with NEPLA-4's spectrum being nearly identical to that of COLA-1 due to the similar size of the near-zone, and that for A370p\_z1 being slightly smaller ($R_{\alpha, \rm{A370p\_z1}} = 0.26$ pMpc). We predict the intrinsic 0.5-7 keV emission from the AGN to be $10^{44.17}$ and $10^{44.09}$ ergs s$^{-1}$ respectively for these LAEs. 

The lensed galaxy MACS1149-JD1 observed at $z \sim 9.11$  \citep{hashimoto2018}  displays a blue-shifted Lyman-$\alpha$ line relative to its [OIII] 88 $\mu$ m with a velocity offset  $\Delta v_{\alpha} = -450$ km/s. This corresponds to  $R_{\alpha} = 0.37$ pMpc, which was difficult to achieve \citep{mason2020} with the observed faintness of the galaxy ($M_{\rm UV} = -18.5$). 
If the ionization zone of the above galaxy arises due to an AGN, we find that the required ionizing emissivity is
$N_{\rm ion} = 1.34 \times 10^{56}$ s$^{-1}$, which corresponds to $M_{1500} = -24.99$, or an intrinsic X-ray luminosity of $10^{44.73}$ ergs s$^{-1}$ in the 0.5 - 7 keV band. 

\section{Obscuration of the AGN}

Since the broad Lyman-$\alpha$ line of the AGN is not detected in the spectra of the LAEs, the AGN source would have to be obscured by a compact torus in its immediate vicinity. We can estimate the extent of this obscuration by comparing the derived properties of the AGN to similar quasars  at these redshifts. 

\citet{calhau2020} studied the radio and X-ray properties of $\sim 4000$ LAEs in the COSMOS SC4K field over $2 < z < 6$ and found a correlation between the hardness ratio, which determines the level of obscuration, and the X-ray luminosity of X-ray detected LAEs, with the more luminous X-ray LAEs observed to have smaller hardness ratios. Our estimate for the intrinsic X-ray luminosity of COLA-1 places it above the hardness 0 category when compared to the \citet{calhau2020} sample, with properties similar to those of obscured AGN discovered in the COSMOS field by \citet{marchesi2016}.   It was found that  for AGN having log $L_X/\text{ergs/s} > 44.1$ in the 2-10 keV range, obscured sources are two to three times as abundant as unobscured ones at $z \sim 6-6.5$ \citep{marchesi2016}.

We can use the the hardness ratio to estimate the column density of the gas surrounding the AGN, using the properties of obscured quasar samples in the literature \citep[e.g.,][see Fig. 14]{peca2020}. This suggests $N_{\rm H} \sim 10^{23.5}$ cm$^{-2}$ for the obscured AGN, which could lead to a few orders of magnitude reduction in observed flux \citep{peca2020}. {For low-$z$ LAEs,  the velocity offset of the red-peak emission line contains information about the neutral hydrogen column density ($N_{\rm HI}$) of the obscuring material in the vicinity of the galaxy \citep[e.g.,][]{shibuya2014, hashimoto2015}.  For the systems considered here, this leads to about $N_{\rm HI} \sim 10^{17} - 10^{20}$ cm$^{-2}$ contributed by the galaxy, which serves as a lower limit to the required obscuration assuming a non-evolution of the velocity offset-column density relation across cosmic time.} Figure  \ref{fig:cola1} shows the decline in the observed flux of the X-ray spectrum  of COLA-1, for three representative column densities obscuring the AGN ($N_{\rm H} = 10^{22.75}, 10^{23.25}$ and $10^{23.75}$ cm$^{-2}$).

\section{Discussion and observational prospects}
We have studied the possible origin of the double-peaked Lyman-Alpha Emitters (LAEs), such as COLA-1, NEPLA4 and A370p\_z1= at $z > 6.5$ in the proximity zone of obscured AGN, whose ionizing emissivity leads to the emergence of the blue wing. In particular, we combined the predictions for the extent of the Lyman-$\alpha$ near-zone of $z \sim 6$ quasars \citep{bolton2007} with the maximum extent of the near-zone required for the blue peak of LAEs to be visible \citep{mason2020}.  In so doing, we find that the required ionizing emissivity may be supplied by AGN with UV magnitudes $M_{\rm UV} \sim -23$ which, assuming typical AGN spectra, correspond to  intrinsic X-ray luminosities of $\sim 10^{44}$ ergs/s in the 2-10 keV range. These values indicate that by improved X-ray sensitivities in the future, it may be possible to confirm or rule out the AGN scenario. High-resolution imaging by the forthcoming \textit{James Webb Space Telescope} might also test the expected conical geometry of the obscured AGN scenario, especially in lensed systems.

 Due to the observed narrowness of the Lyman-$\alpha$ line, such AGN are  likely obscured by the intervening material, making the double peak visible only through the ionization cones. Estimates based on the correlation between the column density and hardness ratio of similar AGN in the \textit{Chandra} field predict the obscuring column density to be $N_{\rm H} \sim 10^{23.5}$ cm$^{-2}$, which may cause a drop by 1-2 orders of magnitude in the observed X-ray luminosity.
 
While the stellar emission in A370\_pz1 was found to be capable of self-ionizing its own {\sc{HII}} bubble (with an escape fraction of $f_{\rm esc} \gtrsim 0.9$ for a fraction of its measured age of about 50 Myr), it was found that COLA-1 and NEPLA-4 need additional sources of ionization to keep their bubbles ionized given galaxy lifetimes of a few ten Myr \citep{meyer2020}. These findings may lend further support to the AGN scenario. Searching for lines such as  [NeV] 3347, 3427 with the \textit{James Webb Space Telescope} will also test this scenario in the future. 

Radio observations of LAEs at $z \sim 6$ with the LOFAR survey \citep{gloudemans2020} have  provided upper limits of $10^{23.90}$ and $10^{23.94}$ W/Hz respectively at $z \sim 5.7$ and $z \sim 6.6$. Converting  the radio upper limits for COLA-1 (Jorryt Matthee and Jo\~ao Calhau, private communication) to their equivalent flux densities assuming a spectral shape with $\alpha = -0.8$ \citep[][]{calhau2020}, we find similar values ($10^{24.59}$ and $10^{23.91}$ W/Hz at 1.4 GHz and 3 GHz respectively). Given that only about 10\% of AGN are radio loud, these upper limits do not constrain the AGN scenario since these objects could be radio quiet. Infrared  observations (1 - 1000 $\mu m$), such as those with the ALMA \citep[e.g.,][]{fudamoto2020} may also be useful, with the caveat that the decline in  AGN flux at these wavelengths may be difficult to estimate, particularly if it is obscured \citep[e.g.,][]{fritz2006}.

\section*{Acknowledgements}  We thank Max Gronke, Charlotte Mason, Jorryt Matthee, Fabio Pacucci and Daniel Schaerer for useful discussions and insightful comments on the manuscript, {and the referee for a helpful report that improved the presentation}. HP acknowledges support from the Swiss National Science Foundation Ambizione Grant PZ00P2\_179934 ``Probing the Universe: through reionization and beyond". The work of AL was supported in part by Harvard's 
Black Hole Initiative, which is funded by grants from JTF and GBMF.

\def\aj{AJ}                   
\def\araa{ARA\&A}             
\def\apj{ApJ}                 
\def\apjl{ApJ}                
\def\apjs{ApJS}               
\def\ao{Appl.Optics}          
\def\apss{Ap\&SS}             
\def\aap{A\&A}                
\def\aapr{A\&A~Rev.}          
\def\aaps{A\&AS}              
\def\azh{AZh}                 
\def\baas{BAAS}
\def\jcap{JCAP}
\def\jrasc{JRASC}             
\def\memras{MmRAS}
\def\na{New Astronomy}
\def\nat{Nature}
\def\mnras{MNRAS}             
\def\pra{Phys.Rev.A}          
\def\prb{Phys.Rev.B}          
\def\prc{Phys.Rev.C}          
\def\prd{Phys.Rev.D}          
\def\prl{Phys.Rev.Lett}       
\def\pasp{PASP}               
\def\pasj{PASJ}
\def\physrep{Phys. Repts.}
\def\qjras{QJRAS}             
\def\skytel{S\&T}             
\def\solphys{Solar~Phys.}     
\def\sovast{Soviet~Ast.}      
\def\ssr{Space~Sci.Rev.}      
\def\zap{ZAp}                 
\let\astap=\aap
\let\apjlett=\apjl
\let\apjsupp=\apjs

\bibliographystyle{aa} 
\bibliography{mybib} 

\end{document}